\documentclass[12pt,onecolumn]{IEEEtran}

\usepackage{amssymb}
\usepackage{graphicx}
\usepackage{amsmath}
\usepackage[colorlinks,linkcolor=blue,citecolor=red,linktocpage=true]{hyperref}

\usepackage[pagewise]{lineno}
\usepackage{cite}

\begin {document}

 %reelle Zahlen

 %natuerliche Zahlen
\def\bbbf{{\rm I\!F}}

\def\bbbz{{\mathchoice {\hbox{$\sf\textstyle Z\kern-0.4em Z$}}
{\hbox{$\sf\textstyle Z\kern-0.4em Z$}}
{\hbox{$\sf\scriptstyle Z\kern-0.3em Z$}}
{\hbox{$\sf\scriptscriptstyle Z\kern-0.2em Z$}}}}

\newtheorem{theorem}{Theorem}
\newtheorem{lemma}{Lemma}
\newtheorem{corollary}{Corollary}
\newtheorem{remark}{Remark}
\newtheorem{example}{Example}

\newcommand{\Tr}{\textrm{\rm Tr}}

%\newenvironment{proof}{\begin{trivlist}\item[]{\em Proof: }}%
%{\samepage \hfill{\hbox{\rlap{$\sqcap$}$\sqcup$}}\end{trivlist}}

\title{The complete weight enumerator of a subclass of optimal three-weight cyclic codes}

\author{
Gerardo Vega\thanks{G. Vega is with the Direcci\'on General de C\'omputo y de Tecnolog\'{\i}as de Informaci\'on y Comunicaci\'on, Uni\-ver\-si\-dad Nacional Aut\'onoma de 
M\'exico, 04510 M\'exico D.F., MEXICO (e-mail: gerardov@unam.mx).} and F\'elix Hern\'andez\thanks{F. Hern\'andez is a PhD student at the Posgrado en Ciencia e Ingenier\'{\i}a de la Computaci\'on, Universidad Nacional Aut\'onoma de M\'exico, 04510 M\'exico, D.F., MEXICO (e-mail: felixhdz@ciencias.unam.mx). Manuscript partially supported by CONACyT, M\'exico.}
%\thanks{Manuscript partially supported by CONACYT, M\'exico.}
}
\maketitle

%\doublespacing

\begin{abstract} 
A class of optimal three-weight cyclic codes of dimension 3 over any finite field was presented by Vega [Finite Fields Appl., 42 (2016) 23-38]. Shortly thereafter, Heng and Yue [IEEE Trans. Inf. Theory, 62(8) (2016) 4501-4513] generalized this result by presenting several classes of cyclic codes with either optimal three weights or a few weights. On the other hand, a class of optimal five-weight cyclic codes of dimension $4$ over a prime field was recently presented by Li, et al. [Adv. Math. Commun., 13(1) (2019) 137-156]. One of the purposes of this work is to present a more general description for these optimal five-weight cyclic codes, which gives place to an enlarged class of optimal five-weight cyclic codes of dimension 4 over any finite field. As an application of this enlarged class, we present the complete weight enumerator of a subclass of the optimal three-weight cyclic codes over any finite field that were studied by Vega [Finite Fields Appl., 42 (2016) 23-38]. In addition, we study the dual codes in this enlarged class of optimal five-weight cyclic codes, and show that they are cyclic codes of length $q^2-1$, dimension $q^2-5$, and minimum Hamming distance $4$. In fact, through several examples, we see that those parameters are the best known parameters for linear codes. 
\end{abstract}


%%%%%%%%%%%%%%%%%%%%%%%%%%%%%%%%%%%%%%%%%%%%%%%%%%%%%%%%%%%%%%%%%%%%%%%%%%%%%%%%%%%
\section{Introduction}\label{secuno}
The problem of obtaining the weight distribution of a given code is important because it plays a significant role in determining the capabilities of error detection and correction of such a code. For cyclic codes this problem is even more important because this kind of codes possess a rich algebraic structure (they are ideals in the principal ideal ring $\bbbf_{q}[x]/(x^n-1)$, where $n$ is the length of the cyclic codes). In addition, it is known that cyclic codes with few weights have a great practical importance in cryptography and coding theory since they are useful in the design of secret sharing schemes and association schemes (see \cite{Anderson,Calderbank}). Not long ago, a characterization of a class of optimal three-weight cyclic codes of dimension 3, over any finite field $\bbbf_{q}$, was presented in \cite{Vega1}, and shortly thereafter, several classes of cyclic codes with either optimal three weights or a few weights were given in \cite{Heng}, showing that one of these classes can be constructed as a generalization of the sufficient numerical conditions of the characterization given in \cite{Vega1}. On the other hand, a class of optimal five-weight $p$-ary cyclic codes of dimension $4$ was recently presented in \cite{Li-Zhu}.

In this paper we take the direct sum (as vector spaces) of a one-weight cyclic code of dimension $2$, and two different one-weight cyclic codes of dimension $1$ in order to construct an enlarged class of optimal five-weight cyclic codes of length $q^2-1$ and dimension $4$, over any finite field $\bbbf_{q}$, with $q\neq 2$, that generalizes the class of optimal five-weight $p$-ary cyclic codes presented in \cite{Li-Zhu}. The codes in this class are optimal in the sense that their lengths reach the Griesmer lower bound for linear codes. In fact, we explicitly determine the weight distribution for the cyclic codes in this class. As an application of this enlarged class of optimal five-weight cyclic codes, we present the complete weight enumerator of a subclass of the optimal three-weight cyclic codes that were studied in \cite{Vega1}. The complete weight enumerator of a code enumerates the codewords by the number of symbols of each kind contained in each codeword. In fact, if $q>2$, the complete weight enumerator contains much more information than the ordinary weight enumerator. For this reason, the determination of the complete weight enumerators of cyclic codes or linear codes over finite fields has received a great deal of attention in recent years (see for example \cite{Bae}, \cite{Chan}, \cite{Li-Bae}, \cite{Li-Yue}, \cite{Yang1}, \cite{Yang2}, \cite{Yang3}, \cite{Yang4}, and \cite{Zheng}). In this respect, it should be pointed out that, for a prime field $\bbbf_p$, several classes of three-weight linear codes and their complete weight enumerators were recently presented in \cite{Kong}, \cite{YangY}, \cite{Yang3}, \cite{Yang4}, and \cite{Zheng}. In that context, we want to emphasize that the three-weight codes that we are going to present here are not only linear, but also cyclic, optimal and defined over any finite field $\bbbf_q$. 

In addition, we study the dual codes in our enlarged class of optimal five-weight cyclic codes, and show that, except for the binary case, they are cyclic codes of length $q^2-1$, dimension $q^2-5$, and minimum Hamming distance $4$. In fact, through several examples, we see that those parameters are the best known parameters for linear codes. 

This work is organized as follows: In Section \ref{secdos} we fix some notations and recall some definitions, along with some known results to be used in subsequent sections. Section \ref{sectres} is devoted to presenting some preliminary and general results. Particularly, we  study a kind of exponential sums that will be important in order to determine the weights, and their corresponding frequencies, for the class of cyclic codes that we are interested in. This kind of exponential sums are then used in Section \ref{seccuatro} to present an enlarged class of optimal five-weight cyclic codes of dimension $4$ over any finite field, showing at the same time that, except for the binary case, the dual codes in this enlarged class have minimum Hamming distance $4$. Examples of optimal five-weight cyclic codes belonging to this enlarged class, along with their corresponding dual codes, are presented at the end of this section. In Section \ref{seccinco} we use our enlarged class of optimal five-weight cyclic codes over any finite field in order to obtain the complete weight enumerator of a subclass of the optimal three-weight cyclic codes that were studied in \cite{Vega1}. Finally, Section \ref{conclusiones} is devoted to conclusions.

%%%%%%%%%%%%%%%%%%%%%%%%%%%%%%%%%%%%%%%%%%%%%%%%%%%%%%%%%%%%%%%%%%%%%%%%%%%%%%%%%%%
\section{Preliminaries}\label{secdos}

In this section, we recall some definitions and notations along with some known results to be used in subsequent sections.

%%%%%%%%%%%%%%%%%%%%%%%%%%%%%%%%%%%%%%%%%%%%%%%%%%%%%%%%%%%%%%%%%%%%%%%%%%%%%%%%%%%
\subsection{Basic definitions and notation}

First of all we set for this section and for the rest of this work, the following:

\medskip
\noindent
{\bf Notation.} Let $p$ be a prime, $q$ a power of $p$, and $\bbbf_{q}$ the finite field of order $q$. By using $\gamma$ we will denote a fixed primitive element of $\bbbf_{q^2}$. We are going to fix $\delta:=\gamma^{q+1}$, and consequently note that $\delta$ is a fixed primitive element of $\bbbf_{q}$. For any integer $a$, the polynomial $h_a(x) \in \bbbf_{q}[x]$ will denote the {\em minimal polynomial} of $\gamma^{-a}$ (see for example \cite[p. 99]{MacWilliams2}). For integers $a_1,a_2,\cdots,a_l$, such that $h_{a_i}(x)\neq h_{a_j}(x)$ if $1 \leq i \neq j \leq l$, ${\cal C}_{(a_1,a_2,\cdots,a_l)}$ will denote the cyclic code of length $q^2-1$ over $\bbbf_{q}$, whose parity check polynomial is $h_{a_1}(x)h_{a_2}(x)\cdots h_{a_l}(x)$. In addition, for any integer $k\geq 1$, we will denote by ``Tr", the absolute trace mapping from $\bbbf_{q^k}$ to the prime field $\bbbf_p$, and by ``$\mbox{Tr}_{\bbbf_{q^k}/\bbbf_q}$" the trace mapping from $\bbbf_{q^k}$ to $\bbbf_q$.

%%%%%%%%%%%%%%%%%%%%%%%%%%%%%%%%%%%%%%%%%%%%%%%%%%%%%%%%%%%%%%%%%%%%%%%%%%%%%%%%%%%
\subsection{Gauss Sums}

The {\em canonical additive character} of $\bbbf_q$ is defined as follows:

$$\chi(x):=e^{2\pi \sqrt{-1} \:\mbox{Tr}(x)/p} \; , \;\;\;\; \mbox{ for all } x \in \bbbf_{q} \; .$$

On the other hand, if $\langle \alpha \rangle = \bbbf_{q}^*$, then any {\em multiplicative character} of $\bbbf_{q}$ is defined by 

$$\psi_j(\alpha^l):=e^{2\pi \sqrt{-1} \: j l/(q-1)} \; , \;\;\; \mbox{ for }  j,l=0,1, \cdots, q-2 \; .$$

\noindent
Commonly $\psi_0$ is referenced as the {\em trivial multiplicative character}. If $q$ is odd, an important multiplicative character is the so-called {\em quadratic character} which is denoted by $\eta$ and defined by: $\eta(x)=1$ if $x$ is the square of an element of $\bbbf_{q}^*$ and $\eta(x)=-1$ otherwise. For the canonical additive character $\chi$, and for any multiplicative character $\psi\neq \psi_0$, of $\bbbf_{q}$, the following two properties will be useful for us 

\begin{eqnarray}\label{eqOrt}
\sum_{x \in \bbbf_{q}^*} \chi(ax)&=&\left\{ \begin{array}{cl}
		q-1 & \mbox{if $a=0$,} \\
\\
		-1 & \mbox{otherwise.}
			\end{array}
\right . \;, \mbox{ and } \nonumber \\
\sum_{x \in \bbbf_{q}^*} \psi(x)&=&0 \;.
\end{eqnarray}

Now, for any multiplicative character $\psi$ of $\bbbf_{q}$ and for the canonical additive character $\chi$ of $\bbbf_{q}$, the {\em Gaussian sum} $G(\psi,\chi)$ is defined by

$$G(\psi,\chi):=\sum_{c \in \bbbf_{q}^*} \psi(c) \chi(c) \; .$$ 

\noindent
Two useful properties of $G(\psi,\chi)$ are (see for example \cite[Theorems 5.12 and 5.15]{Lidl}): 

\begin{equation}\label{eqNueva}
G(\eta,\bar{\chi}) =  \eta(-1) G(\eta,\chi) \; \; \mbox{ and } \; \; G(\eta,\chi)^2=\eta(-1) q \; .
\end{equation}

\noindent
Another important property of Gaussian sums is the so-called expansion of the restriction of $\chi$ to $\bbbf_{q}^*$ in terms of the multiplicative characters of $\bbbf_{q}$, with Gaussian sums as Fourier coefficients (see for example \cite[p. 195]{Lidl}):

$$\chi(c)=\frac{1}{(q-1)}\sum_{\psi} \psi(c) G(\bar{\psi},\chi) \;, \mbox{ for } c \in \bbbf_{q}^* \; .$$

%%%%%%%%%%%%%%%%%%%%%%%%%%%%%%%%%%%%%%%%%%%%%%%%%%%%%%%%%%%%%%%%%%%%%%%%%%%%%%%%%%%
\subsection{Griesmer lower bound}

When constructing a code, from an economical point of view, it is desirable to obtain an $[n,k,d]$ code ${\cal C}$ over $\bbbf_q$ whose length $n$ is minimal for given values of $k$, $d$ and $q$. A lower bound for the length $n$ in terms of these values is as follows. Let $n_q(k,d)$ be the minimum length $n$ for which an $[n,k,d]$ linear code, over $\bbbf_{q}$, exists. If the values of $q$, $k$ and $d$ are given, then a well-known lower bound (see \cite{Griesmer} and \cite{Solomon}) for $n_q(k,d)$ is:

\begin{theorem}\label{teoGriesmer}
(Griesmer lower bound) With the previous notation,

\[n_q(k,d) \geq \sum_{i=0}^{k-1} \left \lceil \frac{d}{q^i} \right \rceil \; ,\]

\noindent
where $\left \lceil x \right \rceil$ denotes the smallest integer greater than or equal to $x$.
\end{theorem} 

As a consequence of the previous theorem we have:

\begin{lemma}\label{lemauno}
Suppose that ${\cal C}$ is a $[q^2-1,4,q(q-1)-2]$ linear code over $\bbbf_{q}$. Then ${\cal C}$ is an optimal linear code in the sense that its length reaches the lower bound in the previous theorem.
\end{lemma} 

\begin{proof} 
By means of a direct application of the Griesmer lower bound, we have

\begin{eqnarray}
&&\left \lceil \frac{q(q-1)-2}{q^0} \right \rceil + \left \lceil \frac{q(q-1)-2}{q} \right \rceil + \left \lceil \frac{q(q-1)-2}{q^2} \right \rceil + \left \lceil \frac{q(q-1)-2}{q^3} \right \rceil \nonumber \\
&=& [(q-1)q-2]+[q-1]+1+1=q^2-1 \; . \nonumber
\end{eqnarray}
\end{proof}

%%%%%%%%%%%%%%%%%%%%%%%%%%%%%%%%%%%%%%%%%%%%%%%%%%%%%%%%%%%%%%%%%%%%%%%%%%%%%%%%%%%
\subsection{The weight enumerator and the complete weight enumerator of a code}

We recall that the {\em weight enumerator} of a code ${\cal C}$ of length $n$ over a finite field is defined as the polynomial $\sum_{j=0}^{n} A_j z^j$, where $A_j\: (0 \leq j \leq n)$ denote the number of codewords with Hamming weight $j$ in the code ${\cal C}$. The sequence $(A_0, A_1,\cdots, A_n)$ is called the {\em weight distribution} of the code. An $M$-\textit{weight} code is a code such that the cardinality of the set of nonzero weights is $M$. That is, $M=|\{i:A_i\neq0,i=1,2,\ldots,n\}|$. 

In a similar way let ${\cal C}$ be a code of length $n$ over $\bbbf_{q}$. Denote the elements of $\bbbf_{q}$ by $u_0=0$, $u_1,\cdots,u_{q-1}$ in some fixed order. For each codeword $\mathbf{c}=(c_0,c_1,\cdots,c_{n-1})$ in ${\cal C}$ let ${\cal Z}(\mathbf{c})$ be the monomial in the $q$ variables $(z_0,z_1,\cdots,z_{q-1})$ given by 

$${\cal Z}(\mathbf{c}):=z_0^{w_0}z_1^{w_1}\cdots z_{q-1}^{w_{q-1}}\;,$$

\noindent
where the power $w_i$ ($0 \leq i < q$) is the number of components $c_j$ ($0 \leq c_j < n$) of $\mathbf{c}$ that are equal to $u_i$. Denote by $V(n,q)$ the set of all integer vectors ${\vec t}=(t_0,t_1,\cdots,t_{q-1})$ such that $t_i \geq 0$ and $\sum_{i=0}^{q-1} t_i=n$. Then the {\em complete weight enumerator} of ${\cal C}$ (see for example \cite{MacWilliams1} and  \cite[p. 141]{MacWilliams2}) is the polynomial

\begin{equation}\label{eqCWE}
\mbox{CWE}_{\cal C}:=\sum_{\mathbf{c} \in {\cal C}} {\cal Z}(\mathbf{c})=\sum_{{\vec t} \in V(n,q)} A_{\vec t} \:Z^{\vec t} \;,
\end{equation}

\noindent
where $Z^{\vec t}=z_0^{t_0}z_1^{t_1}\cdots z_{q-1}^{t_{q-1}}$ and 

$$A_{\vec t}:=\sharp\{\:\mathbf{c} \in {\cal C} \:|\: {\cal Z}(\mathbf{c})=Z^{\vec t}\:\}\;.$$

\noindent
The sequence $(A_{\vec t})_{{\vec t} \in V(n,q)}$ is called the {\em complete weight distribution} of ${\cal C}$. Obviously it coincides with the weight distribution if $q=2$ and contains much more information if $q > 2$. In fact, the complete weight enumerator has a wide range of applications in many research fields as the information it contains is of vital use in practical applications. For example, as pointed out in \cite{Blake} the complete weight enumerator of Reed-Solomon codes could be helpful in soft decision decoding. As other example, the complete weight enumerator is useful in the computation of the Walsh transform of monomial functions over finite fields \cite{Helleseth}.

%%%%%%%%%%%%%%%%%%%%%%%%%%%%%%%%%%%%%%%%%%%%%%%%%%%%%%%%%%%%%%%%%%%%%%%%%%%%%%%%%%%
\section{A class of exponential sums}\label{sectres}

It is well known that the weight distribution of some cyclic codes can be obtained by means of the evaluation of some exponential sums. The following two lemmas goes along these lines.

\begin{lemma}
Let $\chi'$ and $\chi$ be respectively the canonical additive characters of $\bbbf_{q^2}$ and $\bbbf_{q}$. For any integers $e_1$, $e_2$ and $e_3$, and for all $a,b \in \bbbf_{q}$ and $c \in \bbbf_{q^2}$, consider the sums

\[S_{(e_1,e_2,e_3)}(a,b,c) := \sum_{x \in \bbbf_{q^2}^*} \chi(a x^{(q+1)e_1} + b x^{(q+1)e_2}) \chi'(c x^{e_3}) \; .\]

\noindent
If $c\neq 0$ and $\gcd(q+1,e_3)=1$, then

\[S_{(e_1,e_2,e_3)}(a,b,c)=-{\displaystyle \sum_{z \in \bbbf_{q}^*} \sum_{x \in \bbbf_{q}^*}\chi(z+a x^{e_1}+b x^{e_2}+z^{-1}c^{q+1}x^{e_3})} \; .\]
\end{lemma} 

\begin{proof}
Recalling that $\delta:=\gamma^{q+1}$ we have

$$S_{(e_1,e_2,e_3)}(a,b,c)=\sum_{i=0}^{q-2}\chi(a \delta^{i e_1} + b \delta^{i e_2})\sum_{w \in \gamma^i \langle \gamma^{q-1} \rangle}\chi'(cw^{e_3})$$

\noindent
and, since $\gcd(q+1,e_3)=1$, we have 

\[\sum_{w \in \gamma^i \langle \gamma^{q-1} \rangle}\chi'(cw^{e_3})=\sum_{w \in \gamma^{i e_3} \langle \gamma^{q-1} \rangle}\chi'(cw)=\frac{1}{q-1}\sum_{w \in \bbbf_{q^2}^*}\chi'(c \gamma^{i e_3} w^{q-1}) \; .\]

Let $\widehat{\bbbf}_{q^2}$ and $\widehat{\bbbf}_{q}$ be respectively the multiplicative character groups of $\bbbf_{q^2}$ and $\bbbf_{q}$. Now, if $\mbox{N}$ is the norm mapping from $\bbbf_{q^2}$ to $\bbbf_{q}$ and $H$ is the subgroup of order $q-1$ of $\widehat{\bbbf}_{q^2}$, then note that $H=\{\psi \circ \mbox{N} \: | \: \psi \in \widehat{\bbbf}_{q} \}$ (that is, $H$ is nothing but the ``lift" of $\widehat{\bbbf}_{q}$ to $\widehat{\bbbf}_{q^2}$). Therefore, owing to \cite[Theorem 5.30, p. 217]{Lidl}, we have

\begin{eqnarray}
\sum_{w \in \bbbf_{q^2}^*}\chi'(c \gamma^{i e_3} w^{q-1})&=&\sum_{\psi \in \widehat{\bbbf}_{q}} G_{\bbbf_{q^2}}(\bar{\psi} \circ \mbox{N},\chi') \psi(\mbox{N}(c \gamma^{i e_3})) \nonumber \\
&=& -\sum_{\psi \in \widehat{\bbbf}_{q}} G_{\bbbf_{q}}(\bar{\psi},\chi)^2 \psi(\mbox{N}(c \gamma^{i e_3})) \; , \nonumber
\end{eqnarray}

\noindent
where the last equality arises due to the Davenport-Hasse Theorem (see \cite[Theorem 5.14, p. 197]{Lidl}). In consequence, since $\gamma^{i(q+1)}=\mbox{N}(\gamma^i)=\mbox{N}(\gamma)^i=\delta^{i}$ and $ \langle \delta \rangle=\bbbf_{q}^*$, we have 

\begin{equation}\label{equno}
S_{(e_1,e_2,e_3)}(a,b,c)=-\frac{1}{q-1}\sum_{x \in \bbbf_{q}^*}\chi(a x^{e_1} + b x^{e_2}) \sum_{\psi \in \widehat{\bbbf}_{q}} G_{\bbbf_{q}}(\bar{\psi},\chi)^2 \psi(c^{q+1}x^{e_3}) \; .
\end{equation}

On the other hand, by using the Fourier expansion of the restriction of $\chi$ to $\bbbf_{q}^*$ in terms of the multiplicative characters of $\bbbf_{q}$, we have that for all $x,z \in \bbbf_{q}^*$:

\[\chi(z^{-1}c^{q+1}x^{e_3})=\frac{1}{q-1}\sum_{\psi \in \widehat{\bbbf}_{q}} G_{\bbbf_{q}}(\bar{\psi},\chi) \bar{\psi}(z) \psi(c^{q+1}x^{e_3}) \; ,\]

\noindent
and by multiplying both sides of the preceding equation by $\chi(z)$ and by summing we obtain

\[\sum_{z \in \bbbf_{q}^*}\chi(z+z^{-1}c^{q+1}x^{e_3})=\frac{1}{q-1}\sum_{\psi \in \widehat{\bbbf}_{q}} G_{\bbbf_{q}}(\bar{\psi},\chi)^2 \psi(c^{q+1}x^{e_3}) \; .\]

\noindent
Finally, by substituting the previous equation in (\ref{equno}) we obtain the desired result.
\end{proof}

\begin{lemma}\label{lemados}
With the same notation, consider the character sum of the form: 

$$T_{(e_1,e_2,e_3)}(a,b,c) := \sum_{y \in \bbbf_{q}^*} S_{(e_1,e_2,e_3)}(ya,yb,yc)\;.$$

\noindent
If $c\neq 0$, $\gcd(q+1,e_3)=1$, $\gcd(q-1,e_2-e_1)=1$ and $e_3 \equiv e_1+e_2 \pmod{q-1}$, then

$$T_{{(e_1,e_2,e_3)}}(a,b,c)=\left\{ \begin{array}{cl}
		1  & \mbox{if $a$ or $b$ is zero but not both,} \\
		-q^2+q+1 & \mbox{if $a,b\in \bbbf_q^*$ and $a=\frac{c^{q+1}}{b}$,} \\
		q+1 & \mbox{if $a,b\in \bbbf_q^*$ and $a\neq\frac{c^{q+1}}{b}$.}
			\end{array}
\right . $$
\end{lemma} 

\begin{proof}
Without loss of generality we assume that $b\neq0$. Thus, from the previous lemma, and since $y^{q+1}=y^2$, we have

\begin{equation}\label{eqdos}
T_{(e_1,e_2,e_3)}(a,b,c)=-\sum_{x \in \bbbf_{q}^*}\sum_{z \in \bbbf_{q}^*}\sum_{y \in \bbbf_{q}^*}\chi(z+(a + b x^{e_2-e_1})x^{e_1}y+z^{-1}c^{q+1}x^{e_1+e_2} y^2)\;.
\end{equation}
 
\noindent
Now, suppose that $q$ is even. Then by \cite[Theorem 5.34, p. 218]{Lidl} we know that, for all $\rho_0, \rho_1, \rho_2 \in \bbbf_{q}$, 

\begin{equation}\label{eqtres}
\sum_{y \in \bbbf_{q}}\chi(\rho_0+\rho_1y+\rho_2y^2)=
\left\{ \begin{array}{cl}
		\chi(\rho_0)q & \mbox{ if $\rho_2+\rho_1^2=0$,} \\
		0 & \mbox{ otherwise.}
	\end{array}
\right . 
\end{equation}

\noindent
Note that $(a + b x^{e_2-e_1})^2 x^{2e_1}=z^{-1}c^{q+1}x^{e_1+e_2}$ iff $z=\frac{c^{q+1}x^{e_2-e_1}}{(a+b x^{e_2-e_1})^2}$, with $x^{e_2-e_1} \in \bbbf_{q}^* \setminus \{\frac{a}{b}\}$. But, since $\gcd(q-1,e_2-e_1)=1$, the last equality holds iff $z=\frac{c^{q+1}x}{(a+b x)^2}$, with $x \in \bbbf_{q}^* \setminus \{\frac{a}{b}\}$. Therefore 

\begin{eqnarray}
T_{(e_1,e_2,e_3)}(a,b,c)&=&1-q-\sum_{x \in \bbbf_{q}^*} \sum_{z \in \bbbf_{q}^*} \sum_{y \in \bbbf_{q}}\chi(z+(a + b x^{e_2-e_1})x^{e_1}y+z^{-1}c^{q+1}x^{e_1+e_2} y^2) \;, \nonumber \\
&=&1-q-q\sum_{x \in \bbbf_{q}^* \setminus \{\frac{a}{b}\}} \chi(\frac{c^{q+1}x}{(a+bx)^2}) \; , \nonumber  \\
&=&1-q-q(\sum_{x \in \bbbf_{q} \setminus \{\frac{a}{b}\}} \chi(\frac{c^{q+1}x}{(a+bx)^2})-\bar{\delta}_0(a)) \; , \nonumber  \\
&=&1-q+\bar{\delta}_0(a)q-q\sum_{x \in \bbbf_{q} \setminus \{\frac{a}{b}\}} \chi(\frac{c^{q+1}x}{(a+bx)^2}) \; , \nonumber
\end{eqnarray}

\noindent
where $\bar{\delta}_0(a)=1$ if $a\neq 0$ and $0$ otherwise. By making the variable substitution $(a+bx)^{-1} \mapsto w$, with $x \in \bbbf_{q} \setminus \{\frac{a}{b}\}$ (that is $x=\frac{1+aw}{bw}$, with $w \in \bbbf_{q}^*$), we get

\begin{eqnarray}
T_{(e_1,e_2,e_3)}(a,b,c)&=&1-q+\bar{\delta}_0(a)q-q\sum_{w \in \bbbf_{q}^*} \chi(w c^{q+1}(\frac{1+aw}{b})) \; , \nonumber  \\
&=&1-q+\bar{\delta}_0(a)q-q(\sum_{w \in \bbbf_{q}} \chi(\frac{c^{q+1}}{b}w+\frac{c^{q+1}a}{b}w^2)-1) \; , \nonumber  \\
&=&1+\bar{\delta}_0(a)q-q\sum_{w \in \bbbf_{q}} \chi(\frac{c^{q+1}}{b}w+\frac{c^{q+1}a}{b}w^2) \; . \nonumber
\end{eqnarray}

\noindent
Thus, owing to (\ref{eqtres}), the case $q$ even follows from the fact that

$$\sum_{w \in \bbbf_{q}} \chi(\frac{c^{q+1}}{b}w+\frac{c^{q+1}a}{b}w^2)=\left\{ \begin{array}{cl}
		q & \mbox{if $a=\frac{c^{q+1}}{b}$,} \\
		0 & \mbox{otherwise.}
			\end{array}
\right . $$

Now suppose that $q$ is odd. Then by \cite[Theorem 5.33, p. 218]{Lidl} we know that, for all $\rho_0, \rho_1, \rho_2 \in \bbbf_{q}$ with $\rho_2 \neq 0$, 

\begin{equation}\label{eqcuatro}
\sum_{y \in \bbbf_{q}^*}\chi(\rho_0+\rho_1y+\rho_2y^2)=\chi(\rho_0-\rho_1^2(4\rho_2)^{-1})\eta(\rho_2)G_{\bbbf_{q}}(\eta,\chi)-\chi(\rho_0) \; ,
\end{equation}

\noindent
where $\eta$ is the quadratic character of $\bbbf_{q}$. Therefore, from (\ref{eqdos}), we have that $T_{(e_1,e_2,e_3)}(a,b,c)$ is equal to 

\begin{eqnarray}
&&\!\!\!\!\!\!\!\!\!\!\!\!-\sum_{x \in \bbbf_{q}^*} \sum_{z \in \bbbf_{q}^*}\!\chi(z-(a + b x^{e_2-e_1})^2x^{2e_1}(4z^{-1}c^{q+1}x^{e_1+e_2})^{-1})\eta(z^{-1}c^{q+1}x^{e_1+e_2})G_{\bbbf_{q}}(\eta,\chi)-\chi(z) \; , \nonumber  \\
&=&1-q-G_{\bbbf_{q}}(\eta,\chi)\sum_{x \in \bbbf_{q}^*} \sum_{z \in \bbbf_{q}^*}\!\chi(z-(a + b x^{e_2-e_1})^2(4z^{-1}c^{q+1}x^{e_2-e_1})^{-1})\eta(z^{-1}c^{q+1}x^{e_1+e_2}) \; , \nonumber  \\
&=&1-q-G_{\bbbf_{q}}(\eta,\chi)\sum_{x \in \bbbf_{q}^*} \sum_{z \in \bbbf_{q}^*}\!\chi(z-(a + b x^{e_2-e_1})^2(4z^{-1}c^{q+1}x^{e_2-e_1})^{-1})\eta(z^{-1}c^{q+1}x^{e_2-e_1}) \; , \nonumber
\end{eqnarray}

\noindent
where the last equality arises because $\eta(x^{-2e_1})=1$. But $\gcd(q-1,e_2-e_1)=1$, thus after applying the variable substitution $z^{-1}c^{q+1}x^{e_2-e_1} \mapsto w$ (that is $x^{e_2-e_1}=\frac{wz}{c^{q+1}}$), we have that 

\begin{eqnarray}
T_{(e_1,e_2,e_3)}(a,b,c)&=&1-q-G_{\bbbf_{q}}(\eta,\chi)\sum_{w \in \bbbf_{q}^*} \eta(w) \sum_{z \in \bbbf_{q}^*}\!\chi(z-\frac{1}{4w}(a+\frac{bwz}{c^{q+1}})^2) \; , \nonumber  \\
&=&1-q-G_{\bbbf_{q}}(\eta,\chi)\sum_{w \in \bbbf_{q}^*} \eta(w) \sum_{z \in \bbbf_{q}^*}\!\chi(-\frac{a^2}{4w}+(1-\frac{ab}{2c^{q+1}})z-\frac{b^2wz^2}{4c^{2(q+1)}}) \; . \nonumber
\end{eqnarray}

\noindent
After the application of (\ref{eqcuatro}) again, we get that $T_{(e_1,e_2,e_3)}(a,b,c)$ is equal to

\begin{eqnarray}
1-q-G_{\bbbf_{q}}(\eta,\chi)\sum_{w \in \bbbf_{q}^*} \eta(w) [&&\!\!\!\!\!\!\!\!\!\chi(-\frac{a^2}{4w}-(1-\frac{ab}{2c^{q+1}})^2(-\frac{b^2w}{c^{2(q+1)}})^{-1}) \times \nonumber  \\
&&\!\!\!\!\!\!\!\!\!\eta(-\frac{b^2w}{4c^{2(q+1)}})G_{\bbbf_{q}}(\eta,\chi)-\chi(-\frac{a^2}{4w})\:]\;.  \nonumber
\end{eqnarray}

\noindent
But, by considering (\ref{eqNueva}), $G_{\bbbf_{q}}(\eta,\chi)\eta(w)\eta(-\frac{b^2w}{4c^{2(q+1)}})G_{\bbbf_{q}}(\eta,\chi)=\eta(-1)G_{\bbbf_{q}}(\eta,\chi)^2=q$. On the other hand, note that $\sum_{w \in \bbbf_{q}^*} \eta(w) \chi(0)=0$ (see (\ref{eqOrt})) and if $a\neq 0$, then $\eta(w)=\eta(\frac{a^2}{4w})$. Thus 

\begin{eqnarray}
-G_{\bbbf_{q}}(\eta,\chi)\sum_{w \in \bbbf_{q}^*} \eta(w) (-\chi(-\frac{a^2}{4w}))&=&\bar{\delta}_0(a)G_{\bbbf_{q}}(\eta,\chi)\sum_{w \in \bbbf_{q}^*} \eta(\frac{a^2}{4w})\chi(-\frac{a^2}{4w}) \;,  \nonumber \\
&=&\bar{\delta}_0(a)G_{\bbbf_{q}}(\eta,\chi)G_{\bbbf_{q}}(\eta,\bar{\chi})=\bar{\delta}_0(a)q \;.  \nonumber
\end{eqnarray}

\noindent
Therefore

$$T_{(e_1,e_2,e_3)}(a,b,c)=1-q+\bar{\delta}_0(a)q-q\sum_{w \in \bbbf_{q}^*}\chi(-\frac{a^2}{4w}+(1-\frac{ab}{2c^{q+1}})^2(\frac{c^{2(q+1)}}{b^2 w}))\;.$$

\noindent
Applying the variable substitution $\frac{c^{q+1}}{bw} \mapsto v$ (that is $w=\frac{c^{q+1}}{bv}$), we have

\begin{eqnarray}
T_{(e_1,e_2,e_3)}(a,b,c)&=&1-q+\bar{\delta}_0(a)q-q\sum_{v \in \bbbf_{q}^*}\chi(-\frac{a^2bv}{4c^{q+1}}+(1-\frac{ab}{2c^{q+1}})^2(\frac{vc^{q+1}}{b})) \;,  \nonumber \\
&=&1-q+\bar{\delta}_0(a)q-q\sum_{v \in \bbbf_{q}^*}\chi((\frac{c^{q+1}}{b}-a)v) \;.  \nonumber
\end{eqnarray}

\noindent
Finally, the result follows from the fact that (see (\ref{eqOrt}))

$$\sum_{v \in \bbbf_{q}^*}\chi((\frac{c^{q+1}}{b}-a)v)=\left\{ \begin{array}{cl}
		q-1 & \mbox{if $a=\frac{c^{q+1}}{b}$,} \\
		-1 & \mbox{otherwise.}
			\end{array}
\right . $$
\end{proof}

\begin{corollary}\label{coruno}
Assume the same notation as in the previous lemma. If $\gcd(q+1,e_3)=1$, $\gcd(q-1,e_2-e_1)=1$ and $e_3 \equiv e_1+e_2 \pmod{q-1}$, then the character sum $T_{(e_1,e_2,e_3)}(a,b,c)$ takes six different values according to the following seven cases:

\[T_{(e_1,e_2,e_3)}(a,b,c)=
\left\{ \begin{array}{cll}
		\!\!(q-1)(q^2-1) & \mbox{if $c=a=b=0$,} & \mbox{\rm Case 1,} \\
		\!\!1-q^2     & \mbox{if $c=0$ and $a$ or $b$ is zero but not both,} & \mbox{\rm Case 2,} \\
		\!\!q+1     & \mbox{if $c=0$ and $a,b \in \bbbf_q^*$,} & \mbox{\rm Case 3,} \\
		\!\!1-q     & \mbox{if $c\neq 0$ and $a=b=0$,} & \mbox{\rm Case 4,} \\
		\!\!1  & \mbox{if $c\neq 0$ and $a$ or $b$ is zero but not both,} & \mbox{\rm Case 5,} \\
		\!\!-q^2+q+1 & \mbox{if $c\neq 0$ and $a,b\in \bbbf_q^*$ and $a=\frac{c^{q+1}}{b}$,} & \mbox{\rm Case 6,} \\
		\!\!q+1 & \mbox{if $c\neq 0$ and $a,b\in \bbbf_q^*$ and $a\neq\frac{c^{q+1}}{b}$,} & \mbox{\rm Case 7.}
			\end{array}
\right . \]
\end{corollary}

\begin{proof}
Clearly $T_{(e_1,e_2,e_3)}(0,0,0)=(q-1)(q^2-1)$.

Case 2: Without loss of generality we assume that $a=0$ and $b\neq 0$. Then 

$$T_{(e_1,e_2,e_3)}(0,b,0)=\sum_{x \in \bbbf_{q^2}^*} \sum_{y \in \bbbf_{q}^*}\chi(b x^{(q+1)e_2}y)=-(q^2-1) \; .$$
 
Case 3:

\begin{eqnarray}
T_{(e_1,e_2,e_3)}(a,b,0)&=&\sum_{x \in \bbbf_{q^2}^*} \sum_{y \in \bbbf_{q}^*} \chi(a x^{(q+1)e_1}y+b x^{(q+1)e_2}y) \;, \nonumber \\
&=&(q+1)\sum_{x \in \bbbf_{q}^*} \sum_{y \in \bbbf_{q}^*} \chi(a x^{e_1}y+b x^{e_2}y) \;.\nonumber
\end{eqnarray}

\noindent
Applying the variable substitution $x^{e_1}y \mapsto v$ (that is $y=\frac{v}{x^{e_1}}$), we have

$$T_{(e_1,e_2,e_3)}(a,b,0)=(q+1)\sum_{v \in \bbbf_{q}^*} \chi(a v) \sum_{x \in \bbbf_{q}^*} \chi(b v x^{e_2-e_1}) \;.$$

\noindent
But $\gcd(q-1,e_2-e_1)=1$, then

$$T_{(e_1,e_2,e_3)}(a,b,0)=(q+1)\sum_{v \in \bbbf_{q}^*} \chi(a v) \sum_{x \in \bbbf_{q}^*} \chi(b v x)=(q+1) \;.$$

Case 4:

\begin{eqnarray}
T_{(e_1,e_2,e_3)}(0,0,c)&=&\sum_{x \in \bbbf_{q^2}^*} \sum_{y \in \bbbf_{q}^*} \chi(\mbox{Tr}_{\bbbf_{q^2}/\bbbf_{q}}(c x^{e_3})y) \; \nonumber \\
&=&\sum_{i=0}^q \sum_{y \in \bbbf_{q}^*} \sum_{x \in \bbbf_{q}^*} \chi(x^{e_3}\mbox{Tr}_{\bbbf_{q^2}/\bbbf_{q}}(c \gamma^{i e_3})y) \;,\nonumber
\end{eqnarray}

\noindent
but, since $\gcd(q+1,e_3)=1$, we have

\begin{eqnarray}
T_{(e_1,e_2,e_3)}(0,0,c)&=&\sum_{i=0}^q \sum_{x \in \bbbf_{q}^*} \sum_{y \in \bbbf_{q}^*} \chi(x^{e_3}\mbox{Tr}_{\bbbf_{q^2}/\bbbf_{q}}(c \gamma^{i})y)  \nonumber \\
&=&\sum_{x \in \bbbf_{q}^*} \sum_{y \in \bbbf_{q}^*} \chi(0) + q\sum_{x \in \bbbf_{q}^*} \sum_{y \in \bbbf_{q}^*} \chi(y) \nonumber \\
&=&(q-1)^2 -q(q-1)=-(q-1)   \;.\nonumber
\end{eqnarray}

\noindent
The proof of the remaining cases comes from the previous lemma.
\end{proof}

We now summarize our previous results by means of the following:

\begin{remark}\label{remtab}
By considering the frequency of occurrence of all the cases in the previous corollary, we obtain the value distribution of the character sum: 

$$Z_{(e_1,e_2,e_3)}(a,b,c):=\frac{q^2-1}{q}+\frac{1}{q} T_{(e_1,e_2,e_3)}(a,b,c)\;.$$ 

\noindent
The following table depicts such a value distribution:

\medskip
\begin{center}
Table I \\
{\em\small Value distribution of $Z_{(e_1,e_2,e_3)}(a,b,c)$. Here $\gcd(q+1,e_3)=1$, \\ $\gcd(q-1,e_2-e_1)=1$ and $e_3 \equiv e_1+e_2 \pmod{q-1}$.}
\end{center}
\begin{center}
\begin{tabular}{|c|c|c|} \hline
{\bf Value} & $\;$ {\bf Frequency} $\;$ & {\bf Comes from} \\ \hline \hline
$q^2-1$ & 1 & {\rm Case 1} \\ \hline
$0$ & $2(q-1)$ & {\rm Case 2} \\ \hline
$q+1$ & $(q-1)^2(q^2-q-1)$ & {\rm Cases 3 and 7} \\ \hline
$q-1$ & $q^2-1$ & {\rm Case 4} \\ \hline
$q$ & $2(q^2-1)(q-1)$ & {\rm Case 5} \\ \hline
$1$ & $(q^2-1)(q-1)$ & {\rm Case 6} \\ \hline
\end{tabular}
\end{center}
\end{remark}

%%%%%%%%%%%%%%%%%%%%%%%%%%%%%%%%%%%%%%%%%%%%%%%%%%%%%%%%%%%%%%%%%%%%%%%%%%%%%%%%%%%
\section{Optimal five-weight cyclic codes of dimension $4$ and their duals}\label{seccuatro}

The following result gives a full description for the weight distribution of a class of optimal five-weight cyclic codes of length $q^2-1$, dimension $4$ over any finite field $\bbbf_{q}$, with $q\neq 2$.

\begin{center}
Table II \\
{\em Weight distribution of ${\cal C}_{((q+1)e_1,(q+1)e_2,e_3)}$.}
\end{center}
\begin{center}
\begin{tabular}{|c|c|} \hline
{\bf Weight} & $\;$ {\bf Frequency} $\;$\\ \hline \hline
0 & 1 \\ \hline
$q(q-1)-2$ & $(q-1)^2(q^2-q-1)$ \\ \hline
$q(q-1)-1$ & $2(q^2-1)(q-1)$ \\ \hline
$q(q-1)$ & $q^2-1$ \\ \hline
$q^2-2$ & $(q^2-1)(q-1)$ \\ \hline 
$q^2-1$ & $2(q-1)$ \\ \hline 
\end{tabular}
\end{center}

\begin{theorem}\label{teouno}
For integers $e_1$, $e_2$ and $e_3$, let ${\cal C}_{((q+1)e_1,(q+1)e_2,e_3)}$ be the cyclic code of length $q^2-1$, over $\bbbf_{q}$, whose parity-check polynomial is $h_{(q+1)e_1}(x)h_{(q+1)e_2}(x)h_{e_3}(x)$. Thus, if $q\neq 2$, $\gcd(q+1,e_3)=1$, $\gcd(q-1,e_2-e_1)=1$, and $e_3 \equiv e_1+e_2 \pmod{q-1}$ then 

\begin{enumerate}
\item[{\rm (A)}] $h_{e_3}(x)$ is the parity-check polynomial of a one-weight cyclic code of length $q^2-1$, dimension $2$, whose nonzero weight is $q(q-1)$, while $h_{(q+1)e_1}(x)$ and $h_{(q+1)e_2}(x)$ are the parity-check polynomials of two different one-weight cyclic codes of dimension $1$.
\\ 
\item[{\rm (B)}] ${\cal C}_{((q+1)e_1,(q+1)e_2,e_3)}$ is an optimal five-weight $[q^2-1,4,q(q-1)-2]$ cyclic code over $\bbbf_{q}$, whose weight distribution is given by Table II. In addition, if $A_j^{\perp}$, with $0<j\leq q^2-1$, is the number of words of weight $j$ in the dual code of ${\cal C}_{((q+1)e_1,(q+1)e_2,e_3)}$, then $A_1^{\perp}=A_2^{\perp}=A_3^{\perp}=0$ and 

$$A_4^{\perp}=\frac{(q^2-3)(q-1)^2(q-2)^2(q+2)(q+1)}{24} \; .$$
\end{enumerate}
\end{theorem}

\begin{proof}
Part {\rm (A)}: Since $q\neq 2$, $e_1 \not\equiv e_2 \pmod{q-1}$. Thus, $h_{(q+1)e_1}(x)\neq h_{(q+1)e_2}(x)$. On the other hand, $h_{(q+1)e_1}(x)$, $h_{(q+1)e_2}(x)$, and $h_{e_3}(x)$ are the parity-check polynomials of different irreducible cyclic codes of lengths $\frac{q-1}{\gcd(q-1,e_1)}$, $\frac{q-1}{\gcd(q-1,e_2)}$ and $(q+1)\frac{q-1}{\gcd(q-1,e_3)}$, respectively. Therefore, the proof of this part comes now from \cite[Theorem 7]{Vega2}.

Part {\rm (B)}: Clearly, the cyclic code ${\cal C}_{((q+1)e_1,(q+1)e_2,e_3)}$ has length $q^2-1$ and its dimension is $4$. Now, due to Delsarte's Theorem (\cite{Delsarte}), we have

$${\cal C}_{((q+1)e_1,(q+1)e_2,e_3)}:= \{\mathbf{c}(a,b,c) \:|\: a, b \in \bbbf_{q}, c \in \bbbf_{q^2} \}\;,$$

\noindent
where

$$\mathbf{c}(a,b,c):=(a\gamma^{(q+1) i e_1}+b\gamma^{(q+1) i e_2}+\Tr_{\bbbf_{q^2}/\bbbf_q}(c\gamma^{i e_3}))_{i=0}^{q^2-2}\;.$$

\noindent
Thus, the Hamming weight of any codeword $\mathbf{c}(a,b,c)$, in our cyclic code ${\cal C}_{((q+1)e_1,(q+1)e_2,e_3)}$, will be equal to $q^2-1-Z_{(e_1,e_2,e_3)}(a,b,c)$, where

\begin{equation}\label{eqZ1}
Z_{(e_1,e_2,e_3)}(a,b,c)\!=\!\sharp\{\;0 \leq i < q^2-1\; | \; a\gamma^{(q+1) i e_1}+b\gamma^{(q+1) i e_2}+\Tr_{\bbbf_{q^2}/\bbbf_q}(c\gamma^{i e_3})=0 \: \} \:.
\end{equation}

\noindent
If $\chi'$ and $\chi$ are, respectively, the canonical additive characters of $\bbbf_{q^2}$ and $\bbbf_{q}$, then

\begin{eqnarray}
Z_{(e_1,e_2,e_3)}(a,b,c)&=&\frac{1}{q}\sum_{i=0}^{q^2-2} \sum_{y \in \bbbf_q} \chi(ya \gamma^{(q+1) i e_1} + yb \gamma^{(q+1) i e_2}) \chi'(yc \gamma^{i e_3}) \nonumber \\
&=&\frac{q^2-1}{q}+\frac{1}{q}\sum_{y \in \bbbf_q^*} \sum_{x \in \bbbf_{q^2}^*} \chi(ya x^{(q+1)e_1} + yb x^{(q+1)e_2}) \chi'(yc x^{e_3})  \; , \nonumber
\end{eqnarray}   

\noindent
and, by using the notation of Lemma \ref{lemados}, we have

\begin{equation}\label{eqZ2}
Z_{(e_1,e_2,e_3)}(a,b,c)=\frac{q^2-1}{q}+\frac{1}{q} T_{(e_1,e_2,e_3)}(a,b,c) \; .
\end{equation}

\noindent
\noindent
Consequently, the assertion about the weight distribution of ${\cal C}_{((q+1)e_1,(q+1)e_2,e_3)}$ comes now from Remark \ref{remtab}.  

Lastly, ${\cal C}_{((q+1)e_1,(q+1)e_2,e_3)}$ is an optimal cyclic code, due to Lemma \ref{lemauno}, and the assertion about the first four weights of the dual code of ${\cal C}_{((q+1)e_1,(q+1)e_2,e_3)}$ can now be proved by means of the first five identities of Pless (see for example \cite[pp. 259-260]{Huffman} or \cite{Pless}).
\end{proof}

\begin{remark}\label{rmunoE}
By fixing $q \neq 2$, it is not difficult to see that the number, ${\cal N}_{(e_1,e_2,e_3)}(q)$, of different cyclic codes of the form ${\cal C}_{((q+1)e_1,(q+1)e_2,e_3)}$ that satisfy the conditions in Theorem \ref{teouno} is ${\cal N}_{(e_1,e_2,e_3)}(q)= \frac{(q-1)\phi(q^2-1)}{4}$, where $\phi$ denotes the Euler $\phi$-function. Therefore, Theorem \ref{teouno} enlarges the class of optimal five-weight cyclic codes presented in \cite{Li-Zhu}, not only because it is valid for any finite field $\bbbf_q$, with $q \neq 2$, but also, as shown, covers more cases.
\end{remark}

The following are some examples of Theorem \ref{teouno}.

\begin{example}\label{ejecero}
Let $q=3$ and ${\cal C}_{(0,4,1)}$. Then $e_1=0$, $e_2=e_3=1$. Clearly $\gcd(q+1,e_3)=1$, $\gcd(q-1,e_2-e_1)=1$ and $e_3 \equiv e_1+e_2 \pmod{q-1}$. Thus, by Theorem \ref{teouno}, we can see that ${\cal C}_{(0,4,1)}$ is an optimal five-weight $[8,4,4]$ cyclic code over $\bbbf_3$, whose weight enumerator polynomial is

\begin{equation}\label{eqFinal}
1 + 20z^4 + 32z^5 + 8z^6 + 16z^7 + 4z^8 \;.
\end{equation}

On the other hand, the dual code of ${\cal C}_{(0,4,1)}$ is an $[8,4,4]$ cyclic code over $\bbbf_3$ with $A_1^{\perp}=A_2^{\perp}=A_3^{\perp}=0$ and $A_4^{\perp}=20$. In fact, it is not difficult to see that its weight enumerator polynomial is the same as for ${\cal C}_{(0,4,1)}$. 
\end{example}

\begin{example}\label{ejeuno}
Let $q=4$. Thus, by Theorem \ref{teouno} and Remark \ref{rmunoE}, we can see that the $6$ codes: ${\cal C}_{(0,5,1)}$, ${\cal C}_{(0,10,2)}$, ${\cal C}_{(0,5,7)}$, ${\cal C}_{(0,10,11)}$, ${\cal C}_{(5,10,3)}$, and ${\cal C}_{(5,10,6)}$ are optimal five-weight $[15,4,10]$ cyclic codes over $\bbbf_4$, whose weight enumerator polynomial is

\[1 + 99z^{10} + 90z^{11} + 15z^{12} + 45z^{14} + 6z^{15}\;. \]

On the other hand, the dual code of any of these codes is a $[15,11,4]$ cyclic code over $\bbbf_4$ with $A_1^{\perp}=A_2^{\perp}=A_3^{\perp}=0$ and $A_4^{\perp}=585$. 
\end{example}

\begin{example}\label{ejedos}
Let $q=7$ and ${\cal C}_{(0,8,1)}$. Then $e_1=0$, $e_2=e_3=1$. Clearly $\gcd(q+1,e_3)=1$, $\gcd(q-1,e_2-e_1)=1$ and $e_3 \equiv e_1+e_2 \pmod{q-1}$. Thus, by Theorem \ref{teouno}, we can see that ${\cal C}_{(0,8,1)}$ is an optimal five-weight $[48,4,40]$ cyclic code over $\bbbf_7$, whose weight enumerator polynomial is

%A[40]=1476 A[41]=576 A[42]=48 A[47]=288 A[48]=12
\[1 + 1476z^{40} + 576z^{41} + 48z^{42} + 288z^{47} + 12z^{48}\;. \]

%1 + 40z 1476 + 41z 576 + 42z 48 + 47z 288 + 48z 12

On the other hand, the dual code of ${\cal C}_{(0,8,1)}$ is a $[48,44,4]$ cyclic code over $\bbbf_7$ with $A_1^{\perp}=A_2^{\perp}=A_3^{\perp}=0$ and $A_4^{\perp}=124200$. Here is important to note that the weight enumerator polynomial in \cite[Example 7]{Li-Zhu} is wrong.
\end{example}

\begin{remark}
The dual codes in the previous examples have the same parameters as the best known linear codes, according to the tables of the best known linear codes maintained by Markus Grassl at http://www.codetables.de/.
\end{remark}

%%%%%%%%%%%%%%%%%%%%%%%%%%%%%%%%%%%%%%%%%%%%%%%%%%%%%%%%%%%%%%%%%%%%%%%%%%%%%%%%%%%
\section{The complete weight enumerator of a subclass of optimal three-weight cyclic codes}\label{seccinco}

In this section we use our previous results about our enlarged class of optimal five-weight cyclic codes in order to present the complete weight enumerator of a subclass of the optimal three-weight cyclic codes studied in \cite{Vega1}.

\begin{theorem}\label{teotres}
For any two integers $e_2$ and $e_3$, let ${\cal C}_{((q+1)e_2,e_3)}$ be the cyclic code of length $q^2-1$, over $\bbbf_{q}$, given by the set:

$${\cal C}_{((q+1)e_2,e_3)}:= \{\mathbf{c}(b,c) \:|\: b \in \bbbf_{q}, c \in \bbbf_{q^2} \}\;,$$

\noindent
where

$$\mathbf{c}(b,c):=(b\gamma^{(q+1) i e_2}+\Tr_{\bbbf_{q^2}/\bbbf_q}(c\gamma^{i e_3}))_{i=0}^{q^2-2}\;.$$

\noindent
If $\gcd(e_3,q^2-1)=1$ and $e_3 \equiv e_2 \pmod{q-1}$, then ${\cal C}_{((q+1)e_2,e_3)}$ is an optimal three-weight $[q^2-1,3,q(q-1)-1]$ cyclic code of dimension $3$, whose weight enumerator is 

$$1+(q^2-1)(q-1)z^{q(q-1)-1}+(q^2-1)z^{q(q-1)}+(q-1)z^{q^2-1}\;,$$ 

\noindent
while its complete weight enumerator is

$$z_0^{q^2-1} + (q-1)\prod_{i=1}^{q-1}z_{i}^{q+1}+(q^2-1)z_0^{q-1}\prod_{i=1}^{q-1}z_{i}^{q}+(q^2-1)\sum_{j=1}^{q-1}z_0^{q}z_{j}\!\!\!\prod_{i=1,i\neq j}^{q-1}\!\!\!z_{i}^{q+1}\;.$$
\end{theorem}

\begin{proof}
It is not difficult to see that $\gcd(q-1,2e_2-e_3)=1$ and $\gcd(q+1,e_3)=1$, therefore, due to \cite[Theorem 1]{Vega1}, ${\cal C}_{((q+1)e_2,e_3)}$ belongs to the class of optimal three-weight cyclic codes of dimension $3$, whose weight enumerator is the already mentioned.

Now, let $u_0=0$ and $u_l=\gamma^{(q+1)(l-1)}$, with $l=1,2,\cdots,q-1$. Clearly $\{u_0,u_1,\cdots,u_{q-1}\}=\bbbf_q$. Let $\mathbf{c}(b,c)=(c_0,c_1,\cdots,c_{q^2-2})$, $w_i:=w_i(\mathbf{c}(b,c))=\sharp\{\:0\leq j < q^2-1\; | \; c_j=u_i \:\}$. Then, by definition of complete weight enumerator (see (\ref{eqCWE})), we have 

$$\mbox{CWE}_{{\cal C}_{((q+1)e_2,e_3)}}=\sum_{\mathbf{c}(b,c)\in {\cal C}_{((q+1)e_2,e_3)}} {\cal Z}(\mathbf{c}(b,c))\;.$$

\noindent
Let $\bar{1}=(1,1,\cdots,1)$ be the all-ones vector of length $q^2-1$. Thus, observe that

\begin{eqnarray}
w_i&=&q^2-1-w_{H}(\mathbf{c}(b,c)-u_i\bar{1}) \;, \nonumber \\
&=&q^2-1-w_{H}(-u_i\bar{1}+(b\gamma^{(q+1) i e_2}+\Tr_{\bbbf_{q^2}/\bbbf_q}(c\gamma^{i e_3}))_{i=0}^{q^2-2})  \; , \nonumber  \\
&=&\sharp\{\;0 \leq i < q^2-1\; | \; -u_i\gamma^{(q+1) i 0}+b\gamma^{(q+1) i e_2}+\Tr_{\bbbf_{q^2}/\bbbf_q}(c\gamma^{i e_3})=0 \:\} \;, \nonumber 
\end{eqnarray}  

\noindent
and, according to (\ref{eqZ1}), we have

$$w_i=Z_{(e_1,e_2,e_3)}(-u_i,b,c) \;,$$ 

\noindent
where $e_1=0$, $\gcd(q^2-1,e_3)=1$ and $e_3 \equiv e_2 \pmod{q-1}$. But, if the previous conditions hold, then $\gcd(q+1,e_3)=1$, $\gcd(q-1,e_2-e_1)=1$ and $e_3 \equiv e_1+e_2 \pmod{q-1}$. Therefore, owing to (\ref{eqZ2}) and Remark \ref{remtab}, 

$$w_0=\left\{ \begin{array}{cll}
		\!\!q^2-1 & \mbox{if $c=b=0$,} & \mbox{\rm Case 1,} \\
		\!\!0     & \mbox{if $c=0$ and $b\neq 0$,} & \mbox{\rm Case 2,} \\
		\!\!q-1     & \mbox{if $c\neq 0$ and $b=0$,} & \mbox{\rm Case 4,} \\
		\!\!q     & \mbox{if $c\neq 0$ and $b\neq 0$,} & \mbox{\rm Case 5.}
			\end{array}
\right . $$

\noindent
Whereas if $1\leq i \leq q-1$, we have

$$w_i=\left\{ \begin{array}{cll}
		\!\!0 & \mbox{if $c=b=0$,} & \mbox{\rm Case 2,} \\
		\!\!q+1   & \mbox{if $c=0$ and $b\neq 0$,} & \mbox{\rm Case 3,} \\
		\!\!q     & \mbox{if $c\neq 0$ and $b=0$,} & \mbox{\rm Case 5,} \\
		\!\!1     & \mbox{if $c\neq 0$ and $-u_i=\frac{c^{q+1}}{b}$,} & \mbox{\rm Case 6,} \\
		\!\!q+1   & \mbox{if $c\neq 0$ and $-u_i\neq \frac{c^{q+1}}{b}$,} & \mbox{\rm Case 7.}
			\end{array}
\right . $$

\noindent
But ${\cal Z}(\mathbf{c}(b,c))=z_0^{w_0}z_1^{w_1}\cdots z_{q-1}^{w_{q-1}}$, therefore

$${\cal Z}(\mathbf{c}(b,c))=\left\{ \begin{array}{cl}
		\!\!z_0^{q^2-1} & \mbox{if $c=b=0$,}  \\
		\!\!{\displaystyle \prod_{i=1}^{q-1}z_{i}^{q+1}}   & \mbox{if $c=0$ and $b\neq 0$,}  \\
		\!\!{\displaystyle z_0^{q-1}\prod_{i=1}^{q-1}z_{i}^{q}}     & \mbox{if $c\neq 0$ and $b=0$,}  \\
		\!\!{\displaystyle z_0^{q}z_{1} \!\!\!\prod_{i=1,i\neq 1}^{q-1}\!\!\!z_{i}^{q+1}}     & \mbox{if $c,b\neq 0$ and $-u_1=\frac{c^{q+1}}{b}$,}  \\
%		\!\!{\displaystyle z_0^{q}z_{2} \!\!\!\prod_{i=1,i\neq 2}^{q-1}\!\!\!z_{i}^{q+1}}     & \mbox{if $c,b\neq 0$ and $-u_2=\frac{c^{q+1}}{b}$,}  \\
		   \vdots & \;\;\;\;\;\;\;\;\; \vdots  \\
		\!\!{\displaystyle z_0^{q}z_{q-1} \!\!\!\!\!\prod_{i=1,i\neq q-1}^{q-1}\!\!\!\!\!z_{i}^{q+1}}  & \mbox{if $c,b\neq 0$ and $-u_{q-1}=\frac{c^{q+1}}{b}$.} 
			\end{array}
\right . $$

\noindent
Thus, the result now follows from the fact that $\sharp\{\mathbf{c}(b,c)\in {\cal C}_{((q+1)e_2,e_3)} \: | \: b=c=0\}=1$, $\sharp\{\mathbf{c}(b,c)\in {\cal C}_{((q+1)e_2,e_3)} \: | \: c=0 \mbox{ and } b\neq 0\}=q-1$, $\sharp\{\mathbf{c}(b,c)\in {\cal C}_{((q+1)e_2,e_3)} \: | \: c\neq 0 \mbox{ and } b=0\}=q^2-1$, and

$$\sharp\{\mathbf{c}(b,c)\in {\cal C}_{((q+1)e_2,e_3)} \: | \: c,b\neq 0 \mbox{ and } -u_i=\frac{c^{q+1}}{b} \}=q^2-1\;,$$

\noindent
for $i=1,2,\cdots,q-1$.
\end{proof}

\begin{remark}\label{rmdosE}
By fixing $q$ it is not difficult to see that the number, ${\cal N}_{(e_2,e_3)}(q)$, of different cyclic codes of the form ${\cal C}_{((q+1)e_2,e_3)}$ that satisfy the conditions in the previous theorem is ${\cal N}_{(e_2,e_3)}(q)= \frac{\phi(q^2-1)}{2}$, where $\phi$ denotes the Euler $\phi$-function.
\end{remark}

The following is an example of Theorem \ref{teotres}.

\begin{example}\label{ejetres}
Let $q=4$. Thus, by Theorem \ref{teotres} and Remark \ref{rmdosE}, we can see that the $4$ codes: ${\cal C}_{(5,1)}$, ${\cal C}_{(5,7)}$, ${\cal C}_{(10,2)}$, and ${\cal C}_{(10,11)}$ are optimal three-weight $[15,3,11]$ cyclic codes over $\bbbf_4$, whose complete weight enumerator polynomial is

\begin{equation}\label{eqCWE1}
z_0^{15} + 3z_1^{5}z_2^{5}z_3^{5} + 15z_0^{3}z_1^{4}z_2^{4}z_3^{4} + 15z_0^{4}z_1z_2^{5}z_3^{5} + 15z_0^{4}z_1^{5}z_2z_3^{5} + 15z_0^{4}z_1^{5}z_2^{5}z_3 \;.
\end{equation}
\end{example}

\section{Conclusions}\label{conclusiones}
In this work we studied a particular kind of character sum and found its value distribution (Lemma \ref{lemados}, Corollary \ref{coruno}, and Remark \ref{remtab}). We then used this value distribution in order to determine the weight distribution of an enlarged class of optimal five-weight cyclic codes over any finite field $\bbbf_q$, with $q\neq 2$, (Theorem \ref{teouno}) that generalizes the class of optimal five-weight $p$-ary cyclic codes presented in \cite{Li-Zhu}. The codes in this class are optimal in the sense that their lengths reach the Griesmer lower bound for linear codes. As an application of our enlarged class of optimal five-weight cyclic codes, we determine the complete weight enumerator of a subclass of the optimal three-weight cyclic codes (Theorem \ref{teotres}) that were studied in \cite{Vega1}. In this respect, it should be pointed out that, for a prime field $\bbbf_p$, several classes of three-weight linear codes and their complete weight enumerators were recently presented in \cite{Kong}, \cite{YangY}, \cite{Yang3}, \cite{Yang4}, and \cite{Zheng}. In that context, we want to emphasize that the three-weight codes presented here are not only linear, but also cyclic, optimal and defined over any finite field $\bbbf_q$.

In addition, we studied the dual codes in our enlarged class of optimal five-weight cyclic codes and showed that, except for the binary case, they are cyclic codes of length $q^2-1$, dimension $q^2-5$, and minimum Hamming distance 4. In fact, through several examples, we see that the parameters of these dual codes are the best known parameters for linear codes. 

Let ${\cal C}_{(2,5)}$ be the cyclic code of length $8$, over $\bbbf_{3}$, whose parity-check polynomial is $h_{2}(x)h_{5}(x)$. It is not difficult to see that ${\cal C}_{(2,5)}$ is an $[8,4,4]$ optimal five-weight cyclic code, whose weight enumerator is given by (\ref{eqFinal}). Clearly, the weight enumerator of the code ${\cal C}_{(2,5)}$ is in accordance with Table II. However, since $\deg(h_{2}(x))=\deg(h_{5}(x))=2$, this code does not belong to the class of optimal five-weight cyclic codes in Theorem \ref{teouno}. In fact, ${\cal C}_{(2,5)}$ is the dual code of the code in Example \ref{ejecero}. More importantly, the existence of ${\cal C}_{(2,5)}$ shows that there are optimal five-weight cyclic codes whose weight enumerator is in accordance with Table II, but does not satisfy the conditions in Theorem \ref{teouno}. Therefore, it is important to note that Theorem \ref{teouno} cannot be considered as a characterization in the same sense that \cite[Theorem 1]{Vega1} is. In this regard, it would be interesting to find the necessary and sufficient conditions for a cyclic code, of length $q^2-1$ and dimension $4$, to have the weight distribution shown in Table II.

On the other hand, we want to mention that Theorem \ref{teouno} cannot be generalized to cyclic codes of length $q^k-1$ and dimension $k+2$, with $k>2$. This is because there are no optimal five-weight cyclic codes, over $\bbbf_{3}$, of length $3^4-1$, and dimension $6$, constructed as a direct sum (as vector spaces) of a one-weight cyclic code of dimension $4$ and two different one-weight cyclic codes of dimension $1$.

As shown in \cite[Example 2]{Vega1}, there are exactly $12$ optimal three-weight cyclic codes over $\bbbf_4$ of length $15$ and dimension $3$. On the other hand, as shown in Example \ref{ejetres}, at least $4$ codes, of these $12$, have the same complete weight enumerator given by (\ref{eqCWE1}). More precisely, with the notation of Theorem \ref{teotres}, the codes ${\cal C}_{(0,1)}$ and ${\cal C}_{(5,3)}$ are optimal three-weight cyclic codes over $\bbbf_4$ of length $15$ and dimension $3$, whose complete weight enumerators are, respectively,  

\begin{eqnarray}\label{eqCWE2}
&&z_0^{15} + z_1^{15} + z_2^{15} + z_3^{15} + 15z_0^{3}z_1^{4}z_2^{4}z_3^{4} + 15z_0^{4}z_1^{3}z_2^{4}z_3^{4} + 15z_0^{4}z_1^{4}z_2^{3}z_3^{4} + 15z_0^{4}z_1^{4}z_2^{4}z_3^{3} \; \mbox{ and } \nonumber \\
&&z_0^{15} + 3z_1^{5}z_2^{5}z_3^{5} + 5z_0^{3}z_2^{6}z_3^{6}+ 5z_0^{3}z_1^{6}z_3^{6} + 5z_0^{3}z_1^{6}z_2^{6} + 15z_0^{4}z_1^{5}z_2^{3}z_3^{3} + 15z_0^{4}z_1^{3}z_2^{5}z_3^{3} + 15z_0^{4}z_1^{3}z_2^{3}z_3^{5} \;.
\end{eqnarray}  

\noindent
Of course, the above shows that not all the optimal three-weight cyclic codes have the same complete weight enumerator. In fact, in (\ref{eqCWE1}) and (\ref{eqCWE2}) appear the three possible complete weight enumerators for these $12$ codes. Thus, as a complement of this work, we believe that it could be interesting to determine the complete weight enumerator of the remaining part of the optimal three-weight cyclic codes that are beyond Theorem \ref{teotres}. Moreover, we believe that there are exactly $q-1$ possible complete weight enumerators that an optimal three-weight cyclic code, over $\bbbf_q$, can take.

\bibliographystyle{IEEE}

\begin{thebibliography}{1}
\bibitem{Anderson} R. Anderson, C. Ding, T. Helleseth, and T. Kl$\o$ve, ``How to build robust shared control systems," {\it Designs, Codes Cryptogr.}, vol. 15, no. 2, pp. 111-124, 1998.

\bibitem{Bae} S. Bae, C. Li and Q. Yue, ``On the complete weight enumerator of some reducible cyclic codes," {\it Discret. Math.}, vol. 338, pp. 2275-2287, 2015.

\bibitem{Blake} I. F. Blake and K. Kith, ``On the complete weight enumerator of Reed-Solomon codes," {\it SIAM J. Discret. Math.} vol. 4, no. 2, pp. 164-171, 1991.

\bibitem{Calderbank} A. R. Calderbank and J. M. Goethals, ``Three-weight codes and association schemes," {\it Philips J. Res.}, vol. 39, nos. 4-5, pp. 143-152, 1984.

\bibitem{Chan} C. H. Chan and M. Xiong, ``On the Complete Weight Distribution of Subfield Subcodes of Algebraic-Geometric Codes," {\it IEEE Trans. Inf. Theory}, vol. 65, no. 11, pp. 7079-7086, 2019.

\bibitem{Delsarte} P. Delsarte, ``On subfield subcodes of Reed-Solomon codes," {\it IEEE Trans. Inf. Theory}, vol. 21, no. 5, pp. 575-576, 1975.

\bibitem{Griesmer} J. H. Griesmer, ``A bound for error correcting codes," {\it IBM Journal of Res. and Dev.}, vol. 4, no. 5, pp. 532-542, 1960.

\bibitem{Helleseth} T. Helleseth and A. Kholosha, ``Monomial and quadratic bent functions over the finite fields of odd characteristic," {\it IEEE Trans. Inf. Theory}, vol. 52, pp. 2018-2032, 2006.

\bibitem{Heng} Z. Heng and Q. Yue, ``Several classes of cyclic codes with either optimal three weights or a few weights," {\it IEEE Trans. Inf. Theory}, vol. 62, no. 8, pp. 4501-4513, 2016.

\bibitem{Huffman} W. C. Huffman and V. Pless, {\em Fundamentals of Error-Correcting Codes}. Cambridge, U.K.: Cambridge Univ. Press, 2003.

\bibitem{Kong} X. Kong and S. Yang, ``Complete weight enumerators of a class of linear codes with two or three weights,"
{\it Discrete Math.}, vol. 342, no. 11, pp. 3166-3176, 2019.

\bibitem{Li-Bae} C. Li, S. Bae, J. Ahn, S. Yang and Z. A. Yao, ``Complete weight enumerators of some linear codes and their applications," {\it Designs, Codes Cryptogr.}, vol. 81, pp. 153-168, 2016. 

\bibitem{Li-Yue} C. Li, Q. Yue and F. W. Fu, ``Complete weight enumerators of some cyclic codes," {\it Designs, Codes Cryptogr.}, vol. 80, pp. 295-315, 2015.

\bibitem{Li-Zhu} L. Li, S. Zhu and L. Liu, ``The weight distribution of a class of $p$-ary cyclic codes and their applications," {\it Adv. Math. Commun.}, vol. 13, no. 1, pp. 137-156, 2019. 

\bibitem{Lidl} R. Lidl and H. Niederreiter H, {\em Finite Fields}. Cambridge Univ. Press, Cambridge, 1984.

\bibitem{MacWilliams1} F. J. MacWilliams, C. L. Mallows and N. J. A. Sloane, ``Generalizations of Gleason’s theorem on weight enumerators of self-dual codes," {\it IEEE Trans. Inf. Theory}, vol. 18, no. 6, pp. 794-805, 1972.

\bibitem{MacWilliams2} F. J. MacWilliams and N. J. A. Sloane, {\em The Theory of Error-Correcting Codes}. Amsterdam, The Netherlands: North-Holland, 1977.

\bibitem{Pless} V. Pless, ``Power moment identities on weight distributions in error-correcting codes," {\it Inf. Contr.}, vol. 6, pp. 147-152, 1962.

\bibitem{Solomon} G. Solomon and J. J. Stiffler, ``Algebraically punctured cyclic codes," {\it Inform. and Control}, vol. 8, no. 2, pp. 170-179, 1965.

\bibitem{Vega1} G. Vega, ``A characterization of a class of optimal three-weight cyclic codes of dimension 3 over any finite field," {\it Finite Fields Appl.}, vol. 42, pp. 23-38, 2016.

\bibitem{Vega2} G. Vega, ``A characterization of all semiprimitive irreducible cyclic codes in terms of their lengths," {\it Applicable Algebra Eng. Commun. Computing}, vol. 30, no. 5, pp. 441-452, 2019.

\bibitem{Yang1} S. Yang, ``Complete Weight Enumerators of a Class of Linear Codes From Weil Sums," {\it in IEEE Access}, vol. 8, pp. 194631-194639, 2020.

\bibitem{YangY} S. Yang and Z. Yao ``Complete weight enumerators of a family of three-weight linear codes," {\it Des. Codes Cryptogr.}, vol. 82, pp. 663-674, 2017.

\bibitem{Yang2} S. Yang, Z. A. Yao, ``Complete weight enumerators of a class of linear codes," {\it Discret. Math.}, vol. 340, no. 4, pp. 729-739, 2017.

\bibitem{Yang3} S. Yang, Z. A. Yao and C. A. Zhao, ``A class of three-weight linear codes and their complete weight enumerators," {\it Cryptogr. Commun.} vol. 9, pp. 133-149, 2017. 

\bibitem{Yang4} C. Zhu and Q. Liao, ``Complete weight enumerators for several classes of two-weight and three-weight linear codes," {\it Finite Fields Appl.}, vol. 75, 2021, https://doi.org/10.1016/j.ffa.2021.101897.

\bibitem{Zheng} D. Zheng, Q. Zhao, X. Wang and Y. Zhang, ``A class of two or three weights linear codes and their complete weight enumerators," {\it Discrete Math.}, vol. 344, 2021, https://doi.org/10.1016/j.disc.2021.112355.

\end{thebibliography}

\end{document}